\documentclass{svmult}

\usepackage{natbib}
\usepackage{makeidx}     
\usepackage{graphicx}    
\usepackage{multicol}    

\bibpunct{(}{)}{;}{a}{}{,}

\newcommand{\ignore}[1]{}

\newcommand{\ccm}{\mathrm{cm}^{-3}}
\newcommand{\rj}{R_\mathrm{j}}

\makeindex             

\begin{document}

\title*{Large Scale Simulations of Jets \\in Dense and Magnetised Environments}
\titlerunning{Large Scale Simulations of Jets}
\author{Martin Krause
\and
Max Camenzind}
\institute{Landessternwarte K\"onigstuhl
69117 Heidelberg, Germany
\texttt{M.Krause@lsw.uni-heidelberg.de}}
%
%
\maketitle

We have used the vectorised and parallelised
MHD code NIRVANA on the NEC SX-5 in parallel mode
to simulate the interaction of jets with a dense 
environment on a scale of more than 200 jet radii.
A maximum performance of 0.75 GFLOP per processor could be reached.

One simulation is axisymmetric and purely hydrodynamic, but with a
resolution of 20 points per beam-radius (ppb). The bipolar jet is injected in the center
of a spherically symmetric King profile, initially underdense to its environment 
by a factor of 10,000. As expected from our previous work, the jet starts with 
producing a spherical bubble around it, bounded by the bow shock.
The bubble slowly elongates, first with roughly elliptical shape, and then forms
narrower extensions in beam direction. The final aspect ratio of the bow shock is 
1.8. We have transformed the results on a 3D-rectangular grid and integrated 
the emission properties to compare the results with observed central cluster
radio galaxies. In the particular case of Cygnus~A, we come to convincing
consistency, morphologically, regarding the size of the influenced region by the jet,
size, and cylindrical shape of the radio cocoon, and source age.
This strongly supports our earlier hypothesis on the nature of the jet in Cygnus~A,
and the derived constraints on other jet parameters like a power of 
$8 \times 10^{46}$~erg/s and an age of 27~Myr.
But, the simulation also clearly shows the shortcoming of the model:
The jet's beam is very unstable, reaching the tip of the bow shock only very seldom.
Also, the contact discontinuity between shocked beam plasma and shocked ambient
gas is quite disrupted by the action of the Kelvin-Helmholtz-instability.
This is not seen in observations, and necessitates the presence of 
dynamically important magnetic fields or an at least moderately relativistic flow,
or both.

The other simulation
was designed to explore the impact of a jet on a randomly magnetised 
environment. A bipolar jet is injected into a King profile on a 3D 
Cartesian grid, with a resolution of 3~ppb. 
This simulation was not successful, because the timestep became too low
after 0.7~Myr The preliminary results show no increase of the magnetic fields reversal 
scale yet, which was expected from an unpublished 2D slab jet simulation, 
but not that early. 

\section{Introduction}
\label{intro}
Several billion years ago, at redshifts in excess of two, the centers of galaxy clusters,
typically hosting already a relaxed elliptical galaxy with an 
old population of stars, were usually equipped with a powerful radio jet.
There are many different lines of evidence for this \citep{Cea01}.
The host galaxies of radio jets are the brightest ones at their redshift.
Also, there have been found dozens of line emitting objects around five high redshift 
radio galaxies, so far \citep{Venea03}. This is a significant overdensity.
Furthermore, 
the space density of galaxy clusters at low redshift agrees with that of high redshift
radio galaxies. Hence it is clear that radio galaxies pinpoint the most massive 
structures in the high redshift universe.

Contrary to the situation at high redshift, 
the brightest cluster galaxies (BCGs) in the local universe are 
generally associated only with weak radio jets. The only exception being Cygnus~A
(Fig.~\ref{mult}a).
This classical double radio galaxy has an outstanding power, only reached by 
sources with redshifts in excess of roughly unity. 
The affected gas surrounding the radio jet can be observed in great detail 
in the X-ray regime by the Chandra satellite.
Therefore, Cygnus~A can serve as a model for the interaction of the jet with 
the intergalactic medium (IGM), for powerful, classical double radio jets.
Much work has been done on the propagation of extragalactic jets
\citep[compare e.g.][ and references therein]{mypap01a,mypap03a}.
Of particular interest for the present study are simulations that include all the external
gas that is affected by action of the jet 
\citep{CHC97,RHB01,RHB02,SBS02,Saxea02,mypap02c,mypap02d,mypap02a,mypap03a,
	Zanea03}.
These simulations did not yet reach the actual size of the jet in Cygnus~A, 
but could be extrapolated to derive a hydrodynamical model of the jet
\citep{myphd02,mypap02d}. In this model, the jet in Cygnus~A resides in a stratified
galaxy cluster atmosphere, with a central density contrast (jet/IGM) of roughly $10^{-4}$.
We have tested this model with a large scale simulation that we report in sect.~\ref{vlj}.
The radio emission of jets has been applied to study the large scale coherence of 
magnetic fields in galaxy clusters \citep[e.g.][]{DBL02}.
We propose the idea that the magnetic field is randomly oriented,
but with roughly the same plasma $\beta$ everywhere, before the jet passes.
We have already shown, that a slab jet with similar parameters as used here,
produces a coherent field structure in the shocked ambient gas (unpublished),
with reversal scales of 10~to~50~$\rj$, when simulated up to a diameter of 400~$\rj$.
In order to check the validity of this result in 3D, we tried to accomplish such a
simulation. The results are described in sect.~\ref{magsec}.

\subsection{Numerics}
\label{num}
For the computations in this contribution, the magneto-hydrodynamic (MHD) code
{\em Nirvana\_CP} was employed. The main part of this code ({\em NIRVANA}) was
written by Udo Ziegler \cite{ZY97}. In that version, it solves the MHD
equations in three dimensions (3D) for density $\rho$, velocity $\vec{v}$,
internal energy $e$, and magnetic field $\vec{B}$:
\begin{eqnarray}
\frac{\partial \rho}{\partial t} + \nabla \cdot \left( \rho \vec{v}\right)&
 = & 0 \label{conti}\\
\frac{\partial \rho \vec{v}}{\partial t} + \nabla \cdot\left( \rho \vec{v}
\vec{v} \right) & = &
- \nabla p - \rho \nabla \Phi+ \frac{1}{4 \pi} \left(\vec{B} \cdot \nabla \right) \vec{B}
-\frac{1}{8 \pi} \nabla \vec{B}^2 \label{mom}\\
\frac{\partial e}{\partial t} + \nabla \cdot \left(e \vec{v} \right) & = &
- p \; \nabla \cdot \vec{v}\label{ie}\\
\frac{\partial \vec{B}}{\partial t} & = &
\nabla \times ( \vec{v} \times \vec{B}) \label{ind}\enspace ,
\end{eqnarray}
where $\Phi$ denotes an external gravitational potential.

NIRVANA can be characterised by the following properties:
\begin{enumerate}
\item explicit Eulerian time--stepping,
\item operator--splitting  formalism for the advection part of the solver,
\item method of characteristics--constraint--transport algorithm to solve
the induction equation and to compute the Lorentz forces;
\item artificial viscosity has been included to dissipate high--frequency noise
and to allow for shock smearing in case the flow becomes supersonic.
\end{enumerate}

The code was vectorised and parallelised by OpenMP like methods,
and successfully tested on the SX-5 \citep{mypap02b}. All the significant loops could be 
vectorised. The number crunching part scales without significant performance loss.
This is also true for the MHD part of the solver. We show a typical profile output 
below (Tables~\ref{tab:1}~and~\ref{tab:2}), for a run without data output, 
which indicates an optimum in vectorisation
and parallelisation efficiency. The average performance in the 2D simulation was 
only 434 cumulative MFLOPS with eight processors, 
probably because in this run, about 500~GB of data 
had to be dumped to the hard disk, which is a serial process.
\begin{figure}[b]
\sidecaption[h]
\raisebox{2cm}{\mbox{\rotatebox{-90}{\includegraphics[height=.4\textwidth]{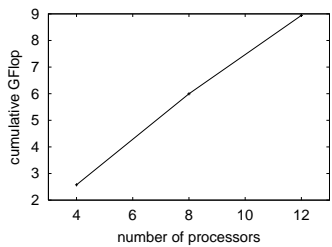}}}}
%
\caption{
Performance for the 3D-MHD problem. The vectorised loops contained 
512 cells, and the parallelised ones 224 cells. With that parameters, the code 
scales very good.}
\label{perf}       
\end{figure}
\begin{table}
\centering
\caption{Typical profile output:  Program Information}
\label{tab:1}       
%
%
\begin{tabular}{l@{:}r@{\hspace{0.5cm}}l@{:}r}
\hline\noalign{\smallskip}
  Real Time (sec)       &           196.546238 &
  User Time (sec)       &          1121.746862 \\
  Sys  Time (sec)       &             5.748649 &
  Vector Time (sec)     &           941.991913 \\
  Inst. Count           &          29681221986 &
  V. Inst. Count        &          12290935628 \\
  V. Element Count      &        3133399958530 &
  FLOP Count            &         851661050792 \\
  MOPS                  &          2808.824657 &
  MFLOPS                &           759.227487 \\
  MOPS   (concurrent)   &         33086.532028 &
  MFLOPS (concurrent)   &          8943.315309 \\
  A.V. Length           &           254.935837 &
  V. Op. Ratio (\%)     &            99.448066 \\
  Memory Size (MB)      &          3216.000000 &
  Max Concurrent Proc.  &                   12 \\
   Conc. Time($>$=1) (sec)&            95.228785 &
   Conc. Time($>$=2) (sec)&            94.162890 \\
   Conc. Time($>$=3) (sec)&            94.106573 &
   Conc. Time($>$=4) (sec)&            94.077965 \\
   Conc. Time($>$=5) (sec)&            94.062470 &
   Conc. Time($>$=6) (sec)&            94.044669 \\
   Conc. Time($>$=7) (sec)&            94.026083 &
   Conc. Time($>$=8) (sec)&            94.009968 \\
   Conc. Time($>$=9) (sec)&            93.985855 &
   Conc. Time($>$=10)(sec)&            93.883754 \\
   Conc. Time($>$=11)(sec)&            93.346376 &
   Conc. Time($>$=12)(sec)&            86.836165 \\
  Event Busy Count      &                    0 &
  Event Wait (sec)      &             0.000000 \\
  Lock Busy Count       &                 1223 &
  Lock Wait (sec)       &             0.945273 \\
  Barrier Busy Count    &                    0 &
  Barrier Wait (sec)    &             0.000000 \\
  MIPS                  &            26.459822 &
  MIPS (concurrent)     &           311.683300 \\
  I-Cache (sec)         &             1.106138 &
  O-Cache (sec)         &            33.629114 \\
  Bank (sec)            &            23.186143 & & \\
\noalign{\smallskip}\hline
\end{tabular}
\end{table}

\begin{table}
\centering
\caption{Typical profile output:  Multitasking Information}
\label{tab:2}       
%
%
\begin{tabular}{rrrr@{}rrrrc}
\hline\noalign{\smallskip}
 \multicolumn{4}{c}{Seconds} & \multicolumn{4}{c}{Seconds} & Thread/Macro[tid]\\
 \%Res. &   Res.  &   T/M  & Micro &  \%CPU  &   CPU& CPUcum. &  Wait &  -micro[n]\\
   0.0  & 93.66  &  0.04  &       &   0.0 &   0.04  &  0.04 &  0.01 & Root1\\
 100.0  &     -  &     -  & 93.62 &   8.4  & 93.62 &  93.66 &  0.77 &  -micro1\\
  99.5  &     -  &     -  & 93.15 &   8.3 &  93.15 & 186.81 &  0.28 &  -micro2\\
  99.9  &     -  &     -  & 93.61 &  8.4  & 93.61 & 280.42  & 0.00  & -micro3\\
  99.5  &     -  &     -  & 93.16 &   8.3 &  93.16 & 373.57 &  0.32 &  -micro4\\
  99.4  &     -   &    -  & 93.11 &   8.3 &  93.11 & 466.68 &  0.22 &  -micro5\\
  99.4  &     -   &    -  & 93.14 &   8.3 &  93.14 & 559.82 &  0.21 &  -micro6\\
  99.3  &     -   &    -  & 93.02 &   8.3 &  93.02 & 652.84 &  0.30 &  -micro7\\
  99.4  &     -  &     -  & 93.13 &   8.3 &  93.13 & 745.97 &  0.31 &  -micro8\\
  99.6  &     -  &     -  & 93.26 &   8.3 &  93.26 & 839.22 &  0.27 &  -micro9\\
  99.5  &     -  &     -  & 93.16 &   8.3  & 93.16 & 932.38 &  0.32 &  -micro10\\
  99.2  &     -  &     -  & 92.93 &   8.3 &  92.93 &1025.31 &  0.26 &  -micro11\\
  99.5  &     -   &    -  & 93.18 &   8.3 &  93.18& 1118.49 &  0.30 &  -micro12\\
\noalign{\smallskip}\hline
\end{tabular}
\begin{tabular}{l@{:}l}
    \%Res.        & The residence ratio.\\
     Res.        & The residence time.\\
     T/M         & The CPU time of thread/macrotask.\\
    Micro        & The CPU time of microtask.\\
    \%CPU         & The CPU ratio of thread/macrotask or microtask.\\
     CPU         & The CPU time of thread/macrotask or microtask.\\
   CPUcum.       & The cumulative CPU time.\\
Thread/Macro[tid]& The thread/macrotask and microtask identifier.\\
\noalign{\smallskip}\hline
\end{tabular}
\end{table}

\section{Simulation of a Very Light Jet to Large Scale}
\label{vlj}
We have performed a bipolar axisymmetric simulation -- in the following called run~A --
of a very light jet in a King type 
galaxy cluster atmosphere with a final jet size of $>200$ jet radii.
The simulation was run for 6200~CPU hours on eight processors of the SX-5 at the HLRS.

\subsection{Simulation Setup}
The jet was injected in both directions (bipolar) 
in the center of a cylindrical grid with 8300 and 2000 points
in axial (Z) and radial (R) direction, respectively. The basic length scale of the 
problem is the jet radius which is represented by 20 points. With that resolution
global parameters like the bow shock velocity on the Z-axis or energy and 
momentum conservation are accurate to $\approx$~10\% \citep{mypap01a}.
As our unit of length we choose the observed jet radius in Cygnus~A, i.e. the jet radius
is set to $\rj=0.5$~kpc. The total grid size is therefore $[207.5\times 50]$~kpc.
The environmental density ($\rho_\mathrm{e}$)
is given by an isothermal King profile:
\begin{equation}\label{kingden}
\rho_\mathrm{e}(R,Z) = \rho_\mathrm{e,0} 
	\left(1+\frac{r^2}{a^2}\right)^{-3\beta/2}\enspace,
\end{equation}
where $r^2=R^2+Z^2$.
This means that the density is constant up to a core radius a that was set to $a=10$~kpc.
Then it starts to decrease, asymptotically reaching $\rho_\mathrm{e}\propto r^{-3\beta}$
($\beta=3/4$). The central density was set to $\rho_\mathrm{e}=m_\mathrm{H}\ccm$,
with the hydrogen mass $m_\mathrm{H}=1.67\times10^{-24}$~g.
The temperature was set to $T=30$~Mio.~K.
This atmosphere is kept in hydrostatic equilibrium by the gravity of a dark matter halo:
\begin{equation}
\Phi=\frac{3\beta k T}{2 \mu m_\mathrm{H}} \log \left(1+\frac{r^2}{a^2}\right)\enspace.
\end{equation}
$\mu$ is the number of particles per proton mass. Here, $\mu=0.5$ for an ionised medium.
In order to brake the symmetry, density perturbations were included, i.e. 
with 10\% probability, the density in a cell was increased by a random factor 
between 0 and 40\%.
The jet density was set to $\rho_\mathrm{j}=\eta_0 \rho_\mathrm{j}$, with the 
density contrast $\eta_0=10^{-4}$. The sound speed in the jet was set to 
$20\% c$, $c$ being the the speed of light, and the jet's Mach number to $M=3$.

The simulation was run for 20~Myr, in total. During that time, the jet reached 
an extention of 110 kpc, i.e. 220 jet radii, on the axis.

\begin{figure}
\vbox{
	\vbox{
		\includegraphics[width=.64\textwidth]{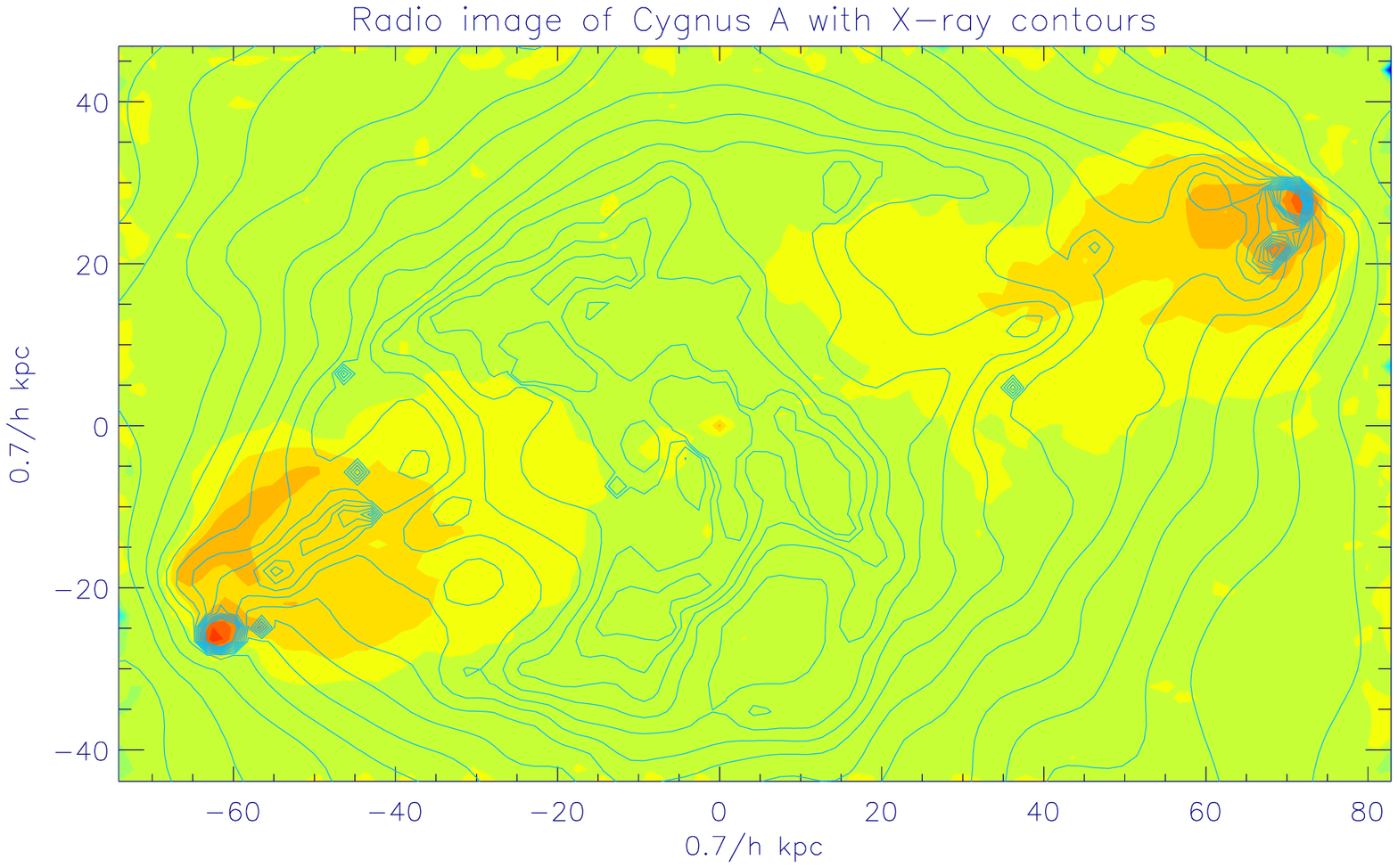}
		\includegraphics[width=.35\textwidth]{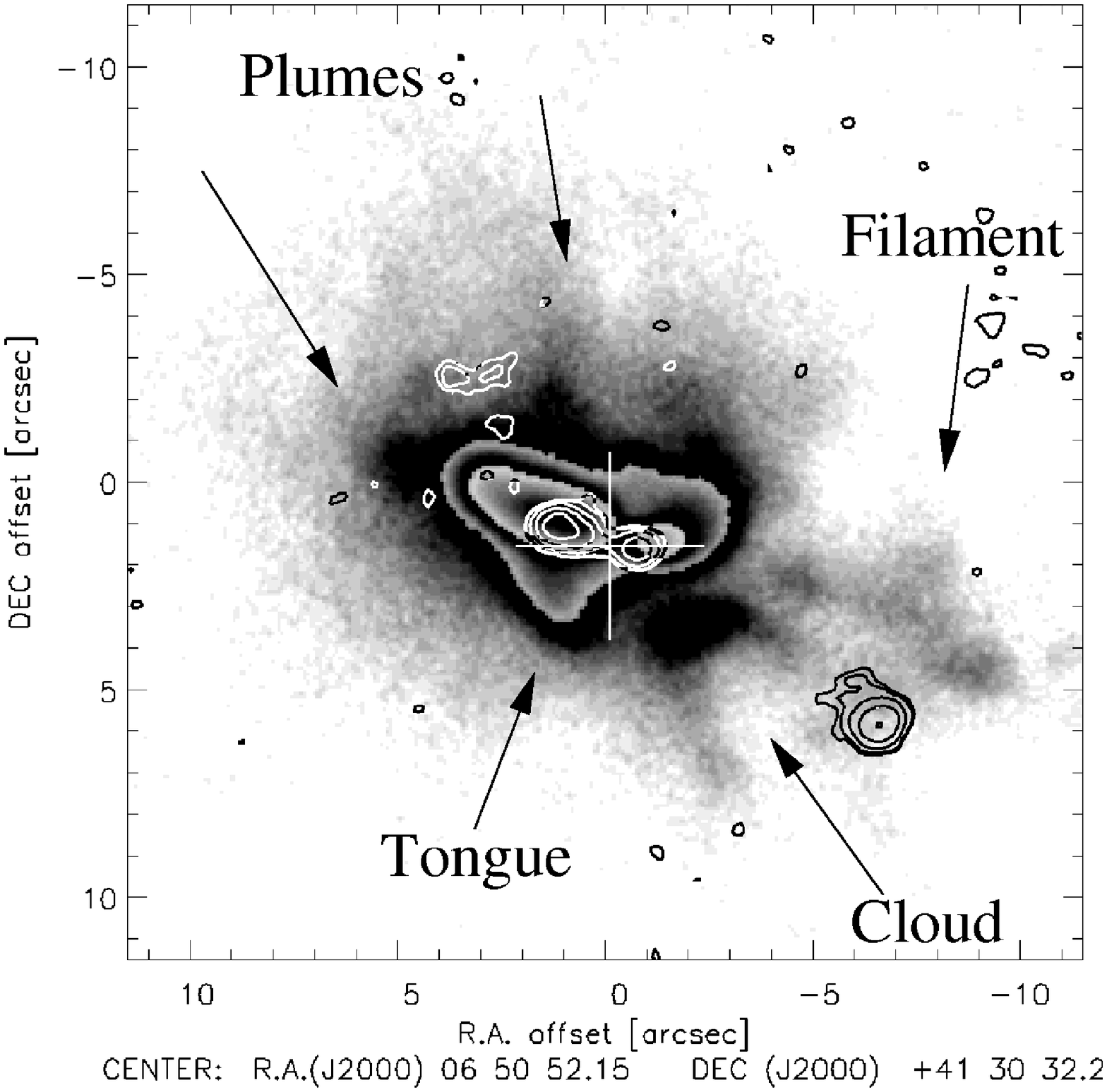}\\
\hfill\mbox{}\hfill\mbox{}\hfill\mbox{(a)}\hfill\mbox{}\hfill\mbox{(b)}\hfill\mbox{}\hfill
		}
	}
\vbox{
	\vbox{
		\rotatebox{-90}{\includegraphics[height=.49\textwidth]{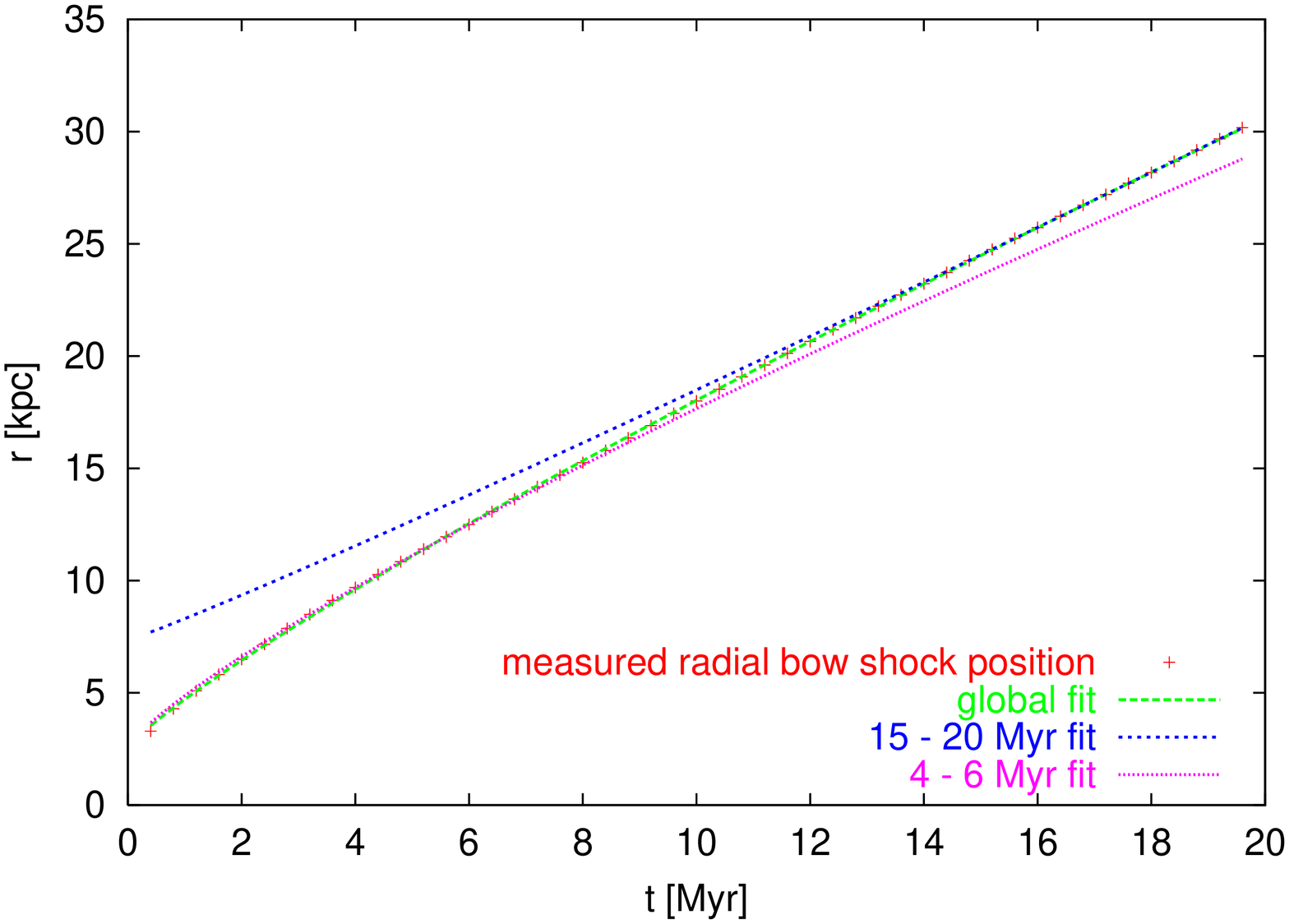}}
		\rotatebox{-90}{\includegraphics[height=.49\textwidth]{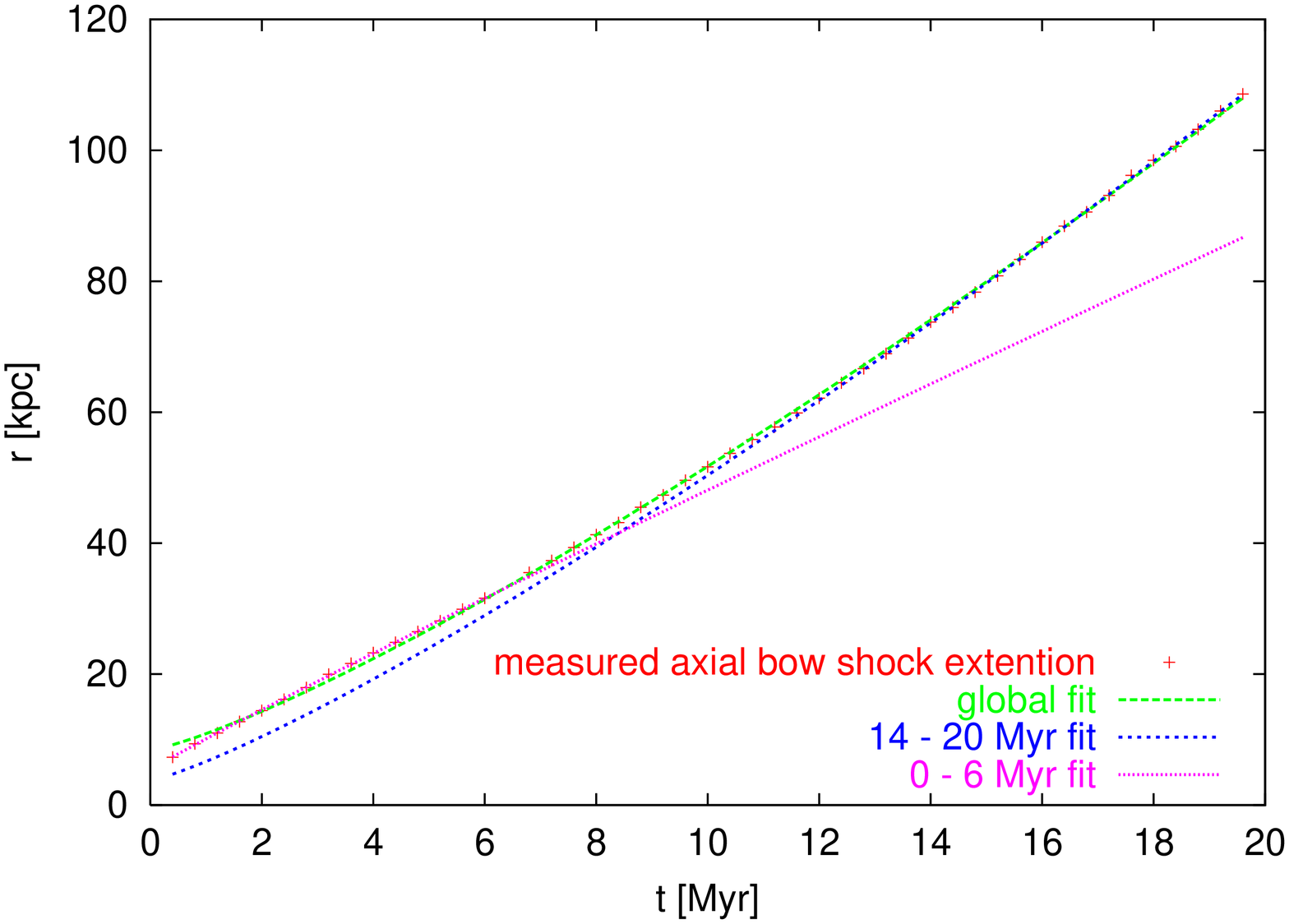}}\\
\hfill\mbox{}\hfill\mbox{(c)}\hfill\mbox{(d)}\hfill\mbox{}\hfill
		}
	}
%
\caption{{\bf (a)}The nearby radio galaxy Cygnus~A. Colours show the logarithmic intensity
	of the 5~GHz radio image (VLA, credits: NRAO / AUI / NSF). Contours show
	the adaptively smoothed 
	X-ray emission (Chandra satellite, credits: NASA / UMD / A.Wilson et al.,
	courtesy: P.~Strub). The axis shows the length scale at the luminosity distance 
	of Cygnus~A (246~$(h/0.7)$~Mpc), where $h$ denotes the Hubble constant in units
	of 100 km/s/Mpc. 
	The jet beam is created at the origin of the coordinate system where 
	the active center of the host galaxy is located. Two barely visible, 
	narrow beams emerge from there in opposite directions, powering the 
	hotspots at (-60,-25) and (70,30). The beam plasma then assembles in a 
	cylindrical cocoon, at lower radio surface brightness. Lower frequency images 
	show that the cocoon continues through the empty region in the center.
	It removes and compresses the IGM, shaping it elliptically.
{\bf (b)}The radio galaxy 4C 41.17 at redshift 3.8, corresponding to a lookback time
of 12 billion years. The gray scale shows the Lyman~$\alpha$
emission line nebula, the contours represent the 5~GHz radio emission.
The white cross indicates the radio core, i.e. the region where usually
the active center of the galaxy is located. 
Adopted from astro-ph: 0303637. Courtesy: Wil van Breugel.
{\bf (c)}Bow shock position at Z=0 versus time for the simulation in 
sect.~\ref{vlj}. The three fits are: 
$2.57017+2.13717 t^{0.859365}$ (global fit),$7.34264+0.960744 t^{1.06464}$ (15-20~Myr),
and $2.59878+2.27161 t^{0.821546}$ (4-6~Myr).
{\bf (d)}Bow shock extention on the axis versus time for the simulation in 
sect.~\ref{vlj}. The three fits are: 
$8.39571+2.51264 t^{1.23664}$ (global fit),$3.73811+2.92394 t^{1.20278}$ (14-20~Myr),
and $5.39429+ 4.71728t^{0.956775}$ (4-6~Myr).
}
\label{mult}       
\end{figure}

\subsection{Results}
\label{res}
\begin{figure}[tb]
\centering
\begin{minipage}{\textwidth}
\begin{center}
\includegraphics[width=.48\textwidth]{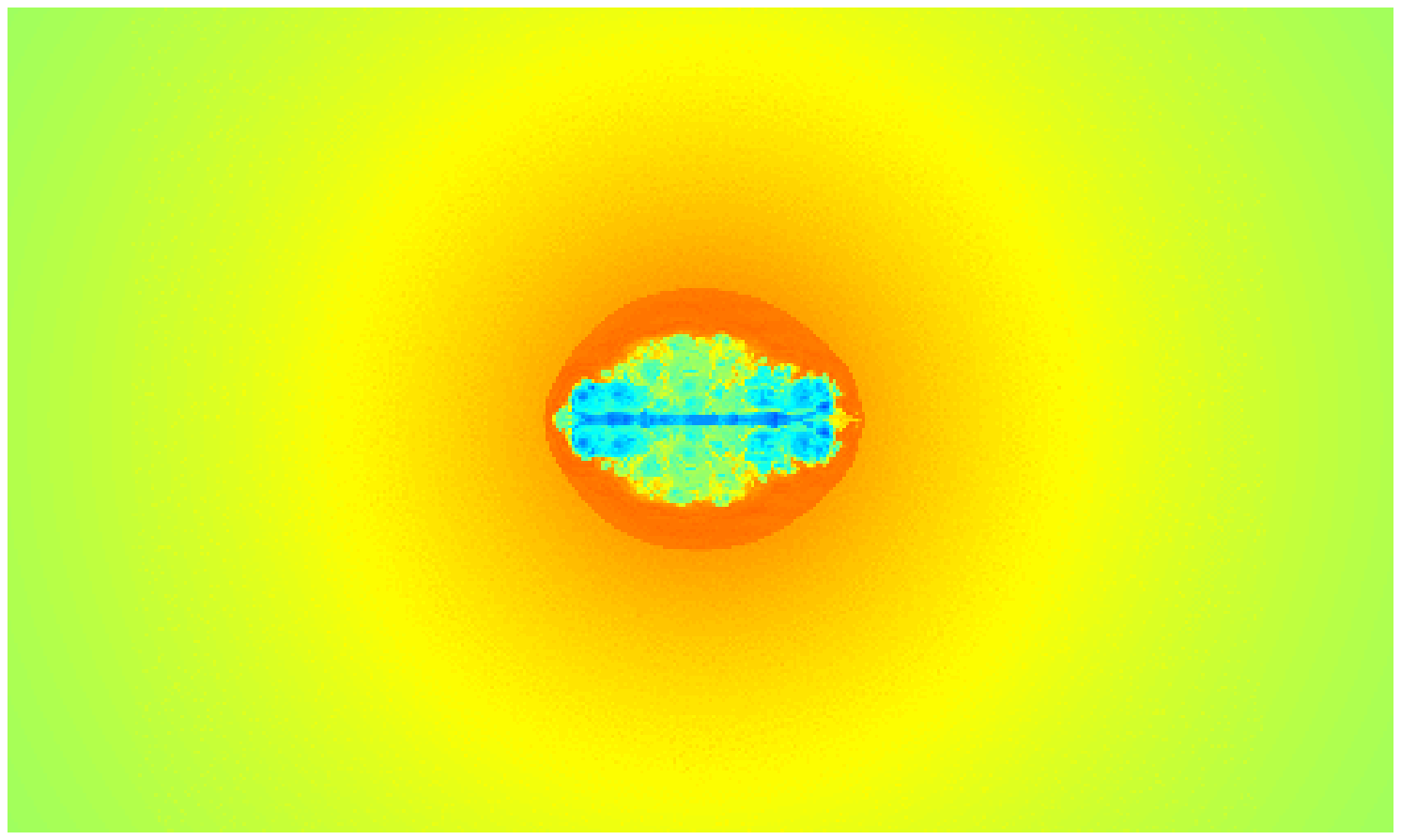}
\includegraphics[width=.48\textwidth]{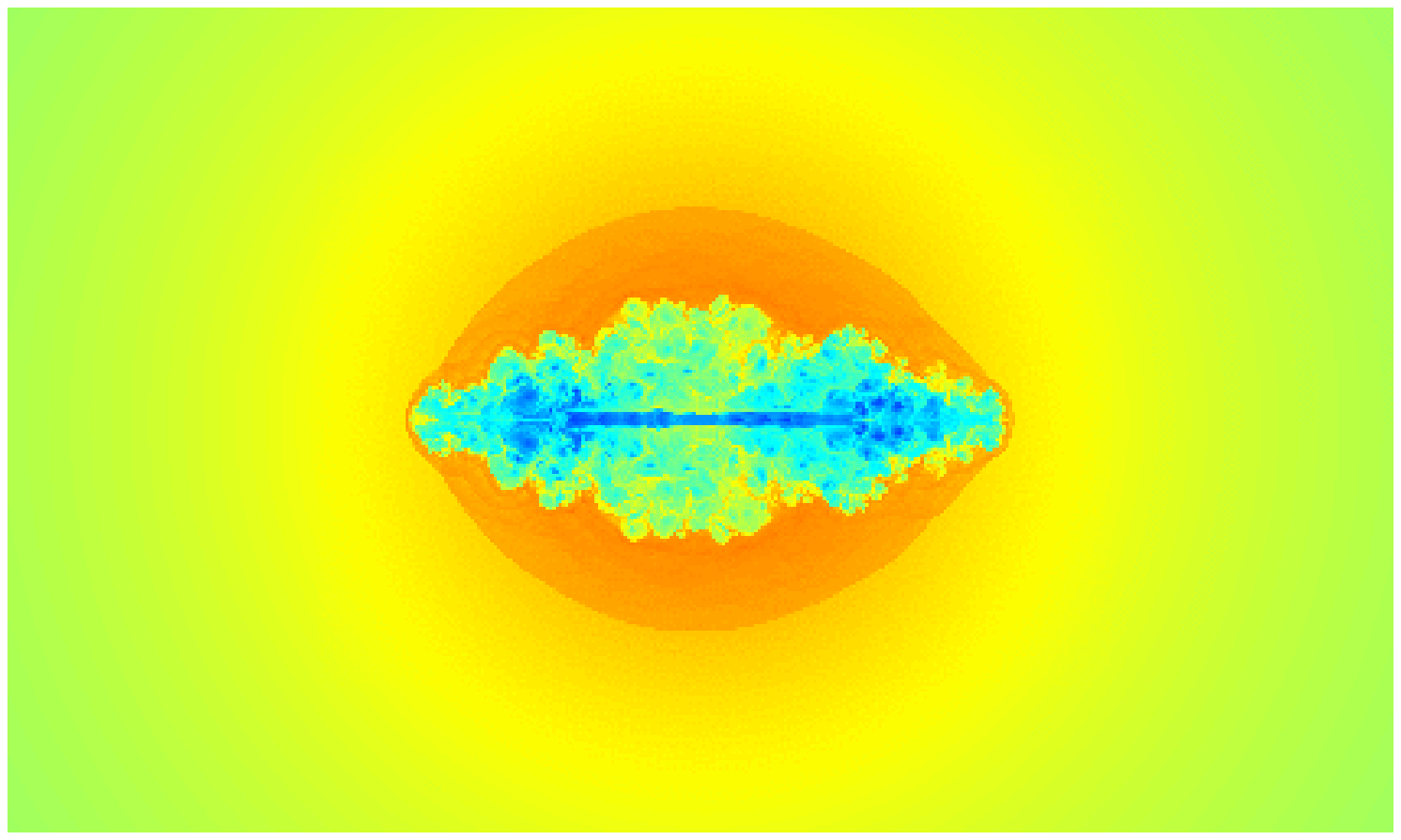}\\
\mbox{}\hfill(a)\hfill\hfill(b)\hfill\mbox{}\\
\includegraphics[width=.48\textwidth]{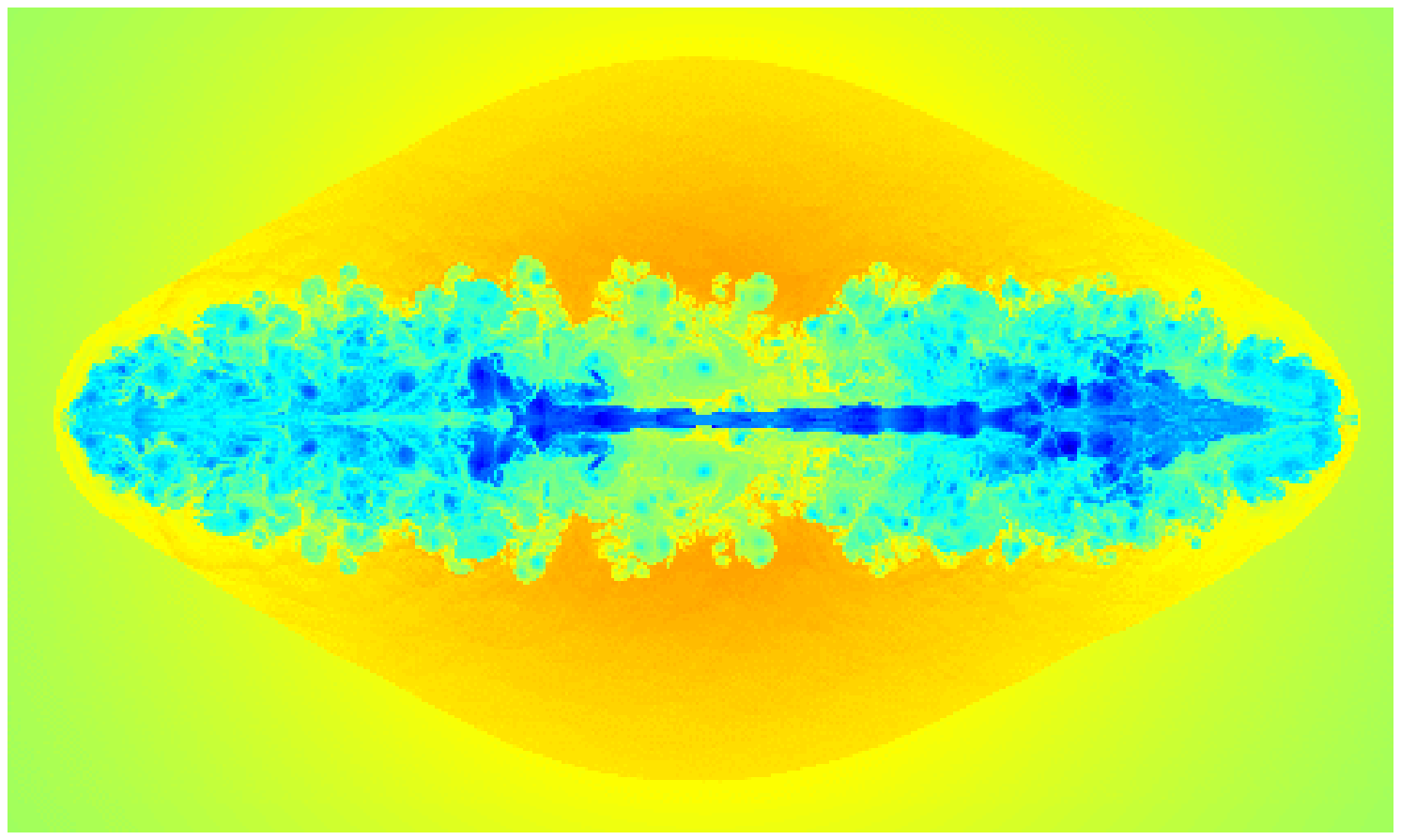}
\includegraphics[width=.48\textwidth]{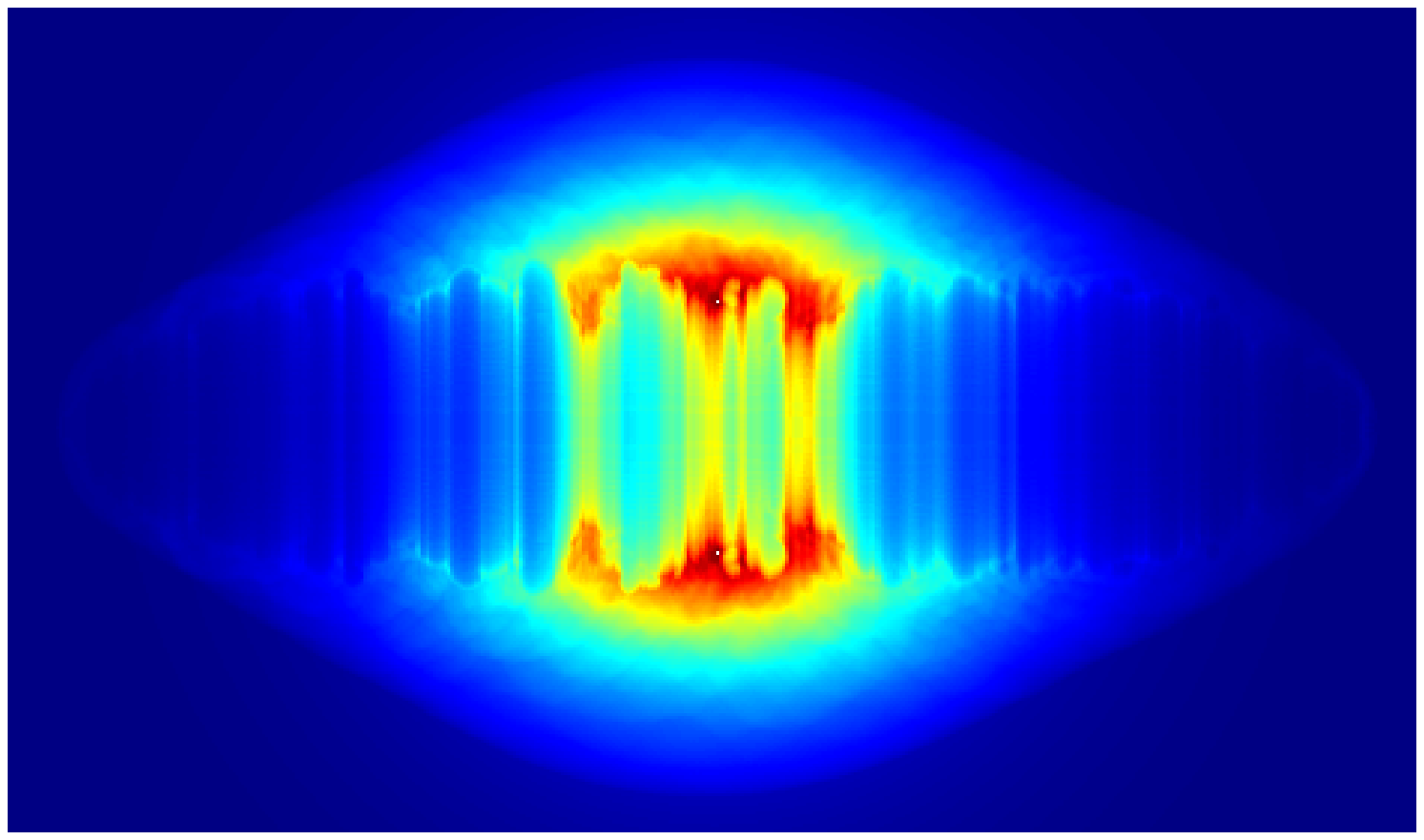}\\
\mbox{}\hfill(c)\hfill\hfill(d)\hfill\mbox{}
\end{center}
\end{minipage}
%
%
\caption{{\bf a-c:} Three 
snapshots of run~A. The logarithm of the density is shown, dark blue
indicates the lowest values, red the highest ones. The times for the snapshots are
5,10,20~Mio. years, for figure a,b, and c, respectively.
The same part of the grid is shown in each case.
The jet forms first a spherical bow shock, associated with a spherical cocoon.
The cocoon then transforms via a conical state to a cylindrical one.
The bow shocks aspect ratio (length/width) also grows, up to 1.8 in c.\newline
{\bf d:} Ray-traced synthetic bremsstrahlung image of (c).}
\label{runAsnaps}       
\end{figure}

We present logarithmic density plots of the simulation results for four 
different simulation times ($5,10,15,20$~Myr) in Fig.~\ref{runAsnaps}.
The morphology that appears in these figures is a continuation of previous 
simulations that could not reach the size shown here.
Fig.~\ref{runAsnaps}a shows the state that was reached by \citet{mypap03a},
and extensively discussed therein. In this early phase, the bow shock is spherical,
its radius following an expansion law given by the force balance equation which can be 
integrated to yield, for arbitrary mass distribution ${\cal M}(r)$ and energy injection
$E(t)$:
\begin{equation}\label{solu}
\int_0^r {\cal M}(r_1) r_1 dr_1 = 2 \int_0^t dt_1 \int_0^{t_1} E(t_2) dt_2.
\end{equation}
For the given matter profile (\ref{kingden}), (\ref{solu}) can be integrated numerically.
We only discuss the asymptotic power law parts here. 
For a power law density distribution ($\rho=\rho_0 (r/r_0)^\kappa$)
and constant energy injection
($E=Lt$), the solution is:
\begin{equation}\label{powsol}
r = \sqrt[\kappa+5]{\frac{(\kappa+3)(\kappa+5)r_0^\kappa L t^3}{12 \pi \rho_0}}
\end{equation}

The density profile used here has the asymptotic power law approximations:
$\lim_{r \mapsto 0}(\rho)=\rho_0$, and $\lim_{r \mapsto \infty}(\rho) \propto r^{-9/4}$.
Therefore, the bow shock should expand with $r\propto t^{0.6}$ in the beginning,
steepening towards $r\propto t^{1.09}$, at least as long as it remains spherical.
The radial bow shock position was determined every 0.4~Myr (compare Fig~\ref{mult}c). 

It was fitted with a function $r=a+bt^c$, where a,b, and c were simultaneously 
varied. At $\approx 5$~Myr, where the bow shock is still almost spherical (aspect 1.2),
$c$ is 0.82. The expected exponent, using a local power law approximation
for the density and the analytic approximation, is:
$c_\mathrm{theo}=0.79$. This confirms the analytic approximation.
But the bow shock continues to follow this expansion law far beyond the spherical phase:
at $\approx 17.6$~Myr, the fit gives $c=1.06$ versus 1.00 from the analytic model.
Even a global power law, with exponent 0.86 fits quite good.

The propagation in axial direction is shown in Fig~\ref{mult}d.
Besides the early times, the bow shock always accelerates.
Its acceleration ($v\propto t^{0.2}$)
is always significantly higher than predicted by self-similar modelling
(asymptotically: $v\propto t^{0.09}$)
\citep[compare e.g.][ and references therein]{CO02a},
but does not reach the prediction for narrow beams of constant radius
that would be super-exponential for $\kappa<-2$.

This reflects the state of the beam plasma.
Figure~\ref{runAsnaps} shows a sound beam, reaching to the tip of the bow shock,
only for $t<10$~Myr. At later phases , the beam is disrupted towards the head.
Thereby, the beam thrust is distributed over a greater area, leading to lower 
bow shock velocity compared to the case of a narrow beam with constant radius.

The cocoon transforms gradually from conical to cylindrical at later times. 
Its width does not grow much for the second half
of the simulation. Towards the center, there is a turbulent region,
where the ambient gas is entrained into the cocoon by the action of gravity, and mixed 
with the shocked beam plasma.

\subsection{Comparison to Observations}
The prime target to compare the simulation data to is the X-ray data of Cygnus~A
(compare Fig.~\ref{mult}a). The simulation was run to almost the size of the 
real source (110~kpc and 140~kpc, for simulation and observation, respectively).
It is evident from Fig.~\ref{runAsnaps} that this is necessary in order to get the 
correct cocoon morphology. Many other details can be understood by this new simulation:
Fig.~\ref{mult}a shows that the radio cocoon is bordered by the brightest X-ray emission. 
We show the line-of-sight integrated X-ray emission for the simulation 
together in Fig.~\ref{runAsnaps}d. 
These can be compared directly to the observations in Fig.~\ref{mult}.
The simulation clearly shows the peaks in emission next to the radio cocoon.
This was already predicted analytically, in self-similar models \citep{Alex02}.
Outwards of this peak, the emissivity is increased with respect to the undisturbed 
cluster emission, by a constant factor in the plane of symmetry. 
The emission falls suddenly at the bow shock.
This detail cannot be seen in the observational data, because there are not enough 
photons (but compare Fig.~4 in \citet{Sea01}). 
Fig.~ \ref{runAsnaps}d shows that the emission of the shocked ambient gas has been shaped 
by the bow shock in an elliptical way.  
\citet{Sea01} have shown that the elliptical isophotal fit is better than the
spherical fit inside a radius of $66/(h/0.7)$~kpc, which should indicate
the bow shock's radial position in Cygnus~A.

The simulation shows significant emission in the mixing region in the central parts of 
the cocoon. However, this is less than observed. The reason is numerical mixing with
cocoon gas in that region. Also, the observation shows dominantly non-axisymmetric modes,
which cannot be represented in our axisymmetric simulation. We note that this region has 
been found to be non-axisymmetric in 3D simulation \citep{mypap02b}.

It has been suggested that repeated jet episodes could heat the surrounding gas
to the observed X-ray temperatures during cosmological timescales.
A lower limit to the heat injection in the cluster gas was derived by \cite{mypap03a},
equ.~(21). For the simulation parameters here, this lower limit amounts to
$3.19\times10^{58}$~erg. The cluster gas was followed by a passive tracer 
that was set to unity. Counting only cells where the tracer variable 
is above 0.1, we measure $8.17\times10^{59}$~erg internal energy injected 
by the jet into the cluster gas by the end of the simulation.
The cluster gas has also gained $2.37\times10^{60}$~erg of potential energy.
In total, 69\% of the energy injected by the jet has been put into the ambient gas.
This is enough to power the X-ray emission of the cluster in Cygnus~A
for $\approx$500~Mio.~years. 
This makes it entirely plausible that the cluster gas is heated 
by that mechanism.

We have proposed to measure jet parameters based on the radial bow shock propagation.
The radial bow shock velocity in Cygnus~A can be constrained from the shock temperature 
measured by Chandra \citep{mypap02d}. This procedure gives a jet power of
$L=8\times10^{46}$~erg/s and an age of $t=27$~Myr for Cygnus~A.
In order to demonstrate the validity of the procedure, we determine the 
jet parameters for the simulation in the same way.
The simplest approximation for the external density is taking it to be constant.
Then, from (\ref{powsol}) it follows with the bow shock velocity of 1220~km/s,
the radial bow shock position 30.17~kpc and the external density of $10^{-25}$~g/cm$^3$:
$L=9.96\times10^{45}$~erg/s and $t=19.7$~Myr. This is to be compared to the true
jet power, $L_\mathrm{true}=\pi \rj^2 \rho_\mathrm{j} v_\mathrm{j}^3=7.72\times10^{45}$~erg/s
and the true jet age of 20~Myr. This very good agreement demonstrates that the exact 
shape of the cluster atmosphere is not critical in determining jet parameters.
The agreement can even be improved by taking better approximations to the density profile.

The simulation result shows an important difference to the observation:
In the simulation, the beam is very unstable, reaching the tip of the bow shock
not even once, after 10~Myr. The reason for this is the low Mach number in the beam
that is a consequence of the strong interaction with the cocoon and the entrained
shocked ambient gas therein. Higher Mach numbers can only be reached 
by a relativistic jet. However, the beam could also be stabilised by an appropriate,
significant magnetic field. The magnetic field is also demanded in order 
to preserve the contact discontinuity near the tip of the bow shock from
the action of the Kelvin-Helmholtz-instability, because the strong 
disruption found in the simulation can not be found in the radio data.
This result is in good agreement with with magnetic field determinations in Cygnus~A's 
hot spots via the self-synchrotron-Compton assumption \citep{WYS00}
and our earlier suggestion, based on a jet power argument, that the jet's
mean Lorentz factor is $\Gamma \approx 20$.

\subsection{Comparison to High Redshift Radio Galaxies}
So far, high redshift radio galaxies have not been observed long enough
in order to study the cluster gas emission in great detail.
But the region where the shocked ambient gas could be expected is typically bright
in emission lines. The emission line regions often have the same cone shaped structure
as the X-ray emission in Cygnus~A (compare Fig~\ref{mult}a~and~\ref{mult}b). 
The line emission is brightest in the region, corresponding to the 
mixing region in the simulation.
This could be interpreted in the way that in these objects the line emission 
is caused by material that was entrained into the 
radio cocoon.\footnote{The alternative interpretation is that cooling is significant 
in the whole shocked ambient gas region. This would lead to very large cocoon width
and turbulent mixing of radio plasma and emission line gas \citep{mypap02a}.
Most recent X-ray data for 4C~41.17 \citep{Scharfea03}, 
showing probably inverse Compton cocoon emission,
point in this second direction for that particular object.}
This gas is subject to the combined thermal and Rayleigh-Taylor instability,
which may cool some gas to the appropriate temperatures \citep{basson02,myphd02}.

\section{A Jet in a Randomly Magnetised Environment}
\label{magsec}
We have accomplished a 3D simulation of a jet in a randomly magnetised environment.
The simulation run for 500~CPU~hours on eight processors of the SX-5.
Unfortunately, the timestep became too low, before significant evolution of the jet 
could be seen.

\subsection{Numerical Setup}
We injected the jet in the center of a Cartesian grid ($[X\times Y\times Z]=
[170\times74\times74]\mathrm{kpc}=[512\times224\times224]$) 
in the X direction. The jet radius was set to$\rj=1$~kpc and resolved with 3 points.
The magnetic field at the jet inlet was set to zero for the $Y$ and the $Z$
direction, and $\partial B_\mathrm{X}/\partial x=0$ on the jet nozzle.
In the surrounding King atmosphere ($\rho_\mathrm{e,0}=m_\mathrm{p}/10$,
$a=10$~kpc , and $\beta=2/3$, $T=3\times10^7$~K), a random magnetic field was established.
The vector potential was randomly determined, folded with the King distribution in order 
to get an average plasma $\beta$ of $\beta=8\pi p/B^2=12$, i.e. sub-equipartition fields.
The initial density contrast was set to $\eta=10^{-3}$ and the Mach number to $M=5$.
\begin{figure}[tbh]
\centering
\begin{minipage}{\textwidth}
\begin{center}
\includegraphics[width=0.49\textwidth]{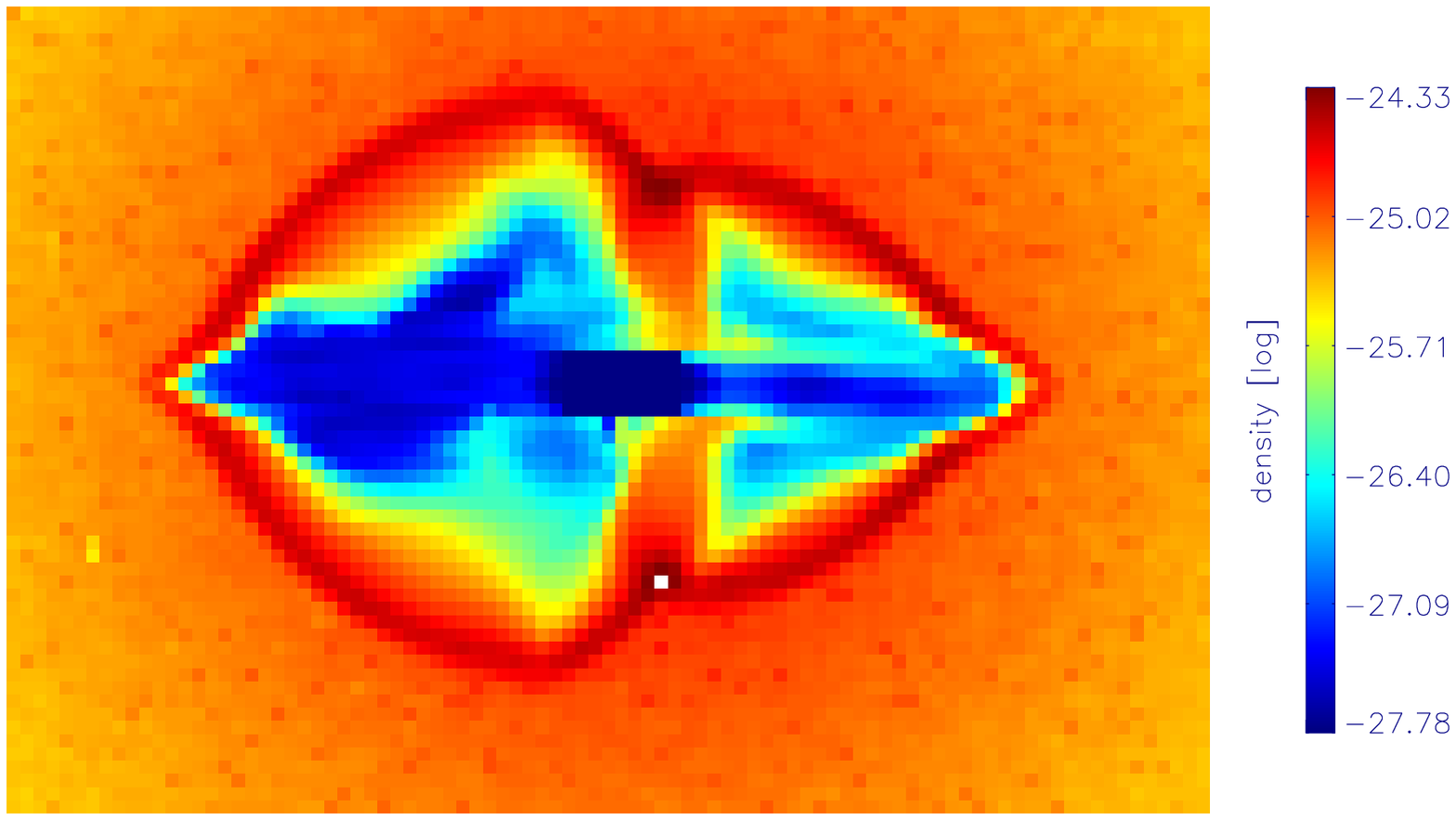}
\includegraphics[width=0.49\textwidth]{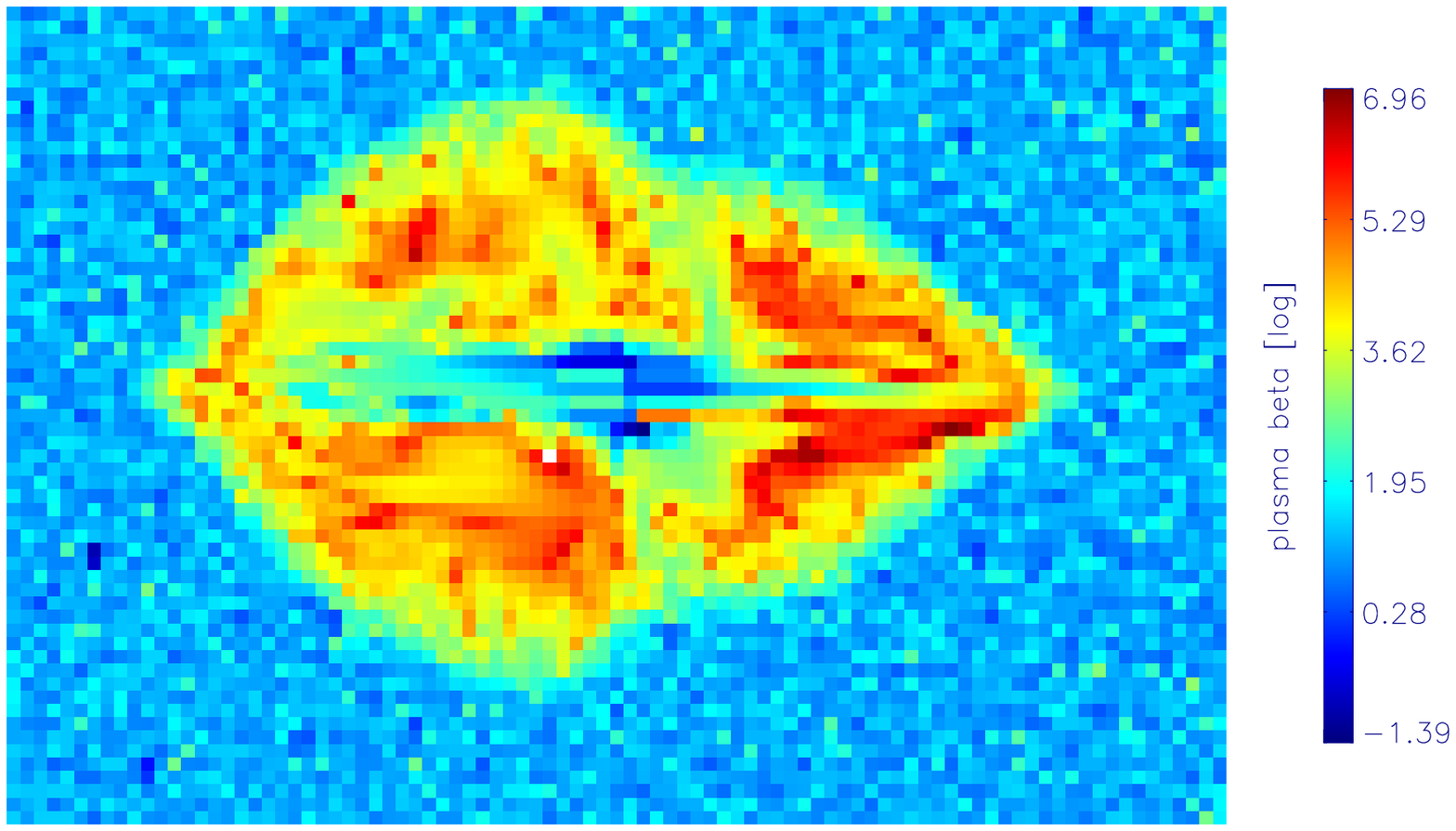}\\
\mbox{}\hfill(a)\hfill\hfill(b)\hfill\mbox{}
\end{center}
\end{minipage}
%
%
\caption{Results of the 3D simulation with randomly magnetised environment at $t=0.7$~Myr.
a: density slice at Z=0, b: plasma beta at Z=0.}
\label{runB}       
\end{figure}

\subsection{Results and Discussion}
A slice of the density distribution and the plasma $\beta$ is shown in Fig.~\ref{runB}.
In that phase, the bow shock is already weak. Weak MHD shocks reduce the field strength 
and refract the field towards the shock normal. 
The simulation shows a decrease of the magnetic field of typically an order of magnitude,
behind the bow shock.
yet. Unfortunately the simulation could not be run for longer time, since the timestep 
became too low. In that phase, no amplification of the reversal scale could be
found, as in a slab simulation (compare sect.~\ref{intro}). This probably happens
only when the shocked ambient gas expands to several times the cocoon diameter,
like e.g. in Fig.~\ref{runAsnaps}d.
If this effect could be confirmed, it would be a good candidate for the explanation 
of the high reversal scale in galaxy clusters, i.e. the coherence of the field is 
created by the weak bow shock.

\section{Conclusions}
Employing the NEC~SX-5 supercomputer at the HLRS, we could simulate 
an axisymmetric jet to the large extention of more than 200 jet radii.
On that scale, the results can be directly compared to observational data of Cygnus~A.
Many details, like elliptically shaped emission, morphology of cocoon
and shocked ambient gas, or low aspect, 
are well reproduced. This confirms the hypothesis that the jet is underdense with 
respect to its environment by a factor of $\approx 10,000$, and the consequences 
on the jet parameters discussed above. We also find differences to the observational data,
concerning stability of the beam and the contact discontinuity, which we ascribe
to the missing magnetic fields and the disregard of relativistic physics.
We speculate that emission line halos of high redshift radio galaxies might be identified
with the centrally concentrated mixing region in the cocoon.

We also tried to simulate the influence of the jet on a randomly magnetised environment.
Unfortunately, the timestep became too low, and the simulation had to be stopped,
before significant progress has been made.

\section*{Acknowledgements}
This work was also supported by the Deutsche Forschungsgemeinschaft 
(Sonderforschungsbereich 439).

\bibliographystyle{apj}
\bibliography{references}

\begin{thebibliography}{23}
\expandafter\ifx\csname natexlab\endcsname\relax\def\natexlab#1{#1}\fi

\bibitem[{{Alexander}(2002)}]{Alex02}
{Alexander}, P. 2002, \mnras, 335, 610

\bibitem[{{Basson}(2002)}]{basson02}
{Basson}, J.~F. 2002, Ph.D.~Thesis, University of Cambridge

\bibitem[{{Carilli} {et~al.}(2001){Carilli}, {Miley}, {R{\"o}ttgering}, {Kurk},
  \& {et al.}}]{Cea01}
{Carilli}, C.~L., {Miley}, G., {R{\"o}ttgering}, H.~J.~A., {Kurk}, J., \& {et
  al.} 2001, in Gas and Galaxy Evolution, ASP Conference Proceedings, Vol. 240.
  eds.: John E. Hibbard, Michael Rupen, and Jacqueline H. van Gorkom

\bibitem[{{Carvalho} \& {O'Dea}(2002)}]{CO02a}
{Carvalho}, J.~C. \& {O'Dea}, C.~P. 2002, \apjs, 141, 337

\bibitem[{{Clarke} {et~al.}(1997){Clarke}, {Harris}, \& {Carilli}}]{CHC97}
{Clarke}, D.~A., {Harris}, D.~E., \& {Carilli}, C.~L. 1997, \mnras, 284, 981

\bibitem[{{Dolag} {et~al.}(2002){Dolag}, {Bartelmann}, \& {Lesch}}]{DBL02}
{Dolag}, K., {Bartelmann}, M., \& {Lesch}, H. 2002, \aap, 387, 383

\bibitem[{{Krause}(2002{\natexlab{a}})}]{mypap02a}
{Krause}, M. 2002{\natexlab{a}}, \aap, 386, L1

\bibitem[{{Krause}(2002{\natexlab{b}})}]{myphd02}
---. 2002{\natexlab{b}}, Ph.D.~Thesis, Universit\"at Heidelberg

\bibitem[{{Krause}(2003)}]{mypap03a}
---. 2003, \aap, 398, 113

\bibitem[{{Krause} \& {Camenzind}(2001)}]{mypap01a}
{Krause}, M. \& {Camenzind}, M. 2001, \aap, 380, 789

\bibitem[{{Krause} \& {Camenzind}(2002{\natexlab{a}})}]{mypap02b}
{Krause}, M. \& {Camenzind}, M. 2002{\natexlab{a}}, in High Performance
  Computing in Science and Engeneering '01, eds.: Krause,~E. and J\"ager,~W.,
  Springer, 329+

\bibitem[{{Krause} \& {Camenzind}(2002{\natexlab{b}})}]{mypap02c}
{Krause}, M. \& {Camenzind}, M. 2002{\natexlab{b}}, in Active Galactic Nuclei:
  from Central Engine to Host Galaxy, meeting held in Meudon, France, July
  23-27, 2002, Eds.: S. Collin, F. Combes and I. Shlosman. To be published in
  ASP Conference Series, p. 42.

\bibitem[{{Krause} \& {Camenzind}(2002{\natexlab{c}})}]{mypap02d}
{Krause}, M. \& {Camenzind}, M. 2002{\natexlab{c}}, in The Physics of
  Relativistic Jets in the CHANDRA and XMM Era , meeting held in Bologna,
  Italy, September 23-27, 2002, Eds.: G. Brunetti, D.E. Harris, R.M. Sambruna
  and G. Setti. To be published in New Astronomy Review

\bibitem[{{Reynolds} {et~al.}(2001){Reynolds}, {Heinz}, \& {Begelman}}]{RHB01}
{Reynolds}, C.~S., {Heinz}, S., \& {Begelman}, M.~C. 2001, \apjl, 549, L179

\bibitem[{{Reynolds} {et~al.}(2002){Reynolds}, {Heinz}, \& {Begelman}}]{RHB02}
---. 2002, \mnras, 332, 271+

\bibitem[{{Saxton} {et~al.}(2002{\natexlab{a}}){Saxton}, {Bicknell}, \&
  {Sutherland}}]{SBS02}
{Saxton}, C.~J., {Bicknell}, G.~V., \& {Sutherland}, R.~S. 2002{\natexlab{a}},
  \apj, 579, 176

\bibitem[{{Saxton} {et~al.}(2002{\natexlab{b}}){Saxton}, {Sutherland},
  {Bicknell}, {Blanchet}, \& {Wagner}}]{Saxea02}
{Saxton}, C.~J., {Sutherland}, R.~S., {Bicknell}, G.~V., {Blanchet}, G.~F., \&
  {Wagner}, S.~J. 2002{\natexlab{b}}, \aap, 393, 765

\bibitem[{{Scharf} {et~al.}(2003){Scharf}, {Smail}, {Ivison}, {Bower}, {van
  Breugel}, \& {Reuland}}]{Scharfea03}
{Scharf}, C.~A., {Smail}, I., {Ivison}, R., {Bower}, R.~G., {van Breugel}, W.,
  \& {Reuland}, M. 2003, ArXiv Astrophysics e-prints astro-ph/0306314

\bibitem[{{Smith} {et~al.}(2002){Smith}, {Wilson}, {Arnaud}, {Terashima}, \&
  {Young}}]{Sea01}
{Smith}, D.~A., {Wilson}, A.~S., {Arnaud}, K.~A., {Terashima}, Y., \& {Young},
  A.~J. 2002, \apj, 565

\bibitem[{{Venemans} {et~al.}(2003){Venemans}, {Miley}, {Kurk}, {Rottgering},
  \& {Pentericci}}]{Venea03}
{Venemans}, B., {Miley}, G., {Kurk}, J., {Rottgering}, H., \& {Pentericci}, L.
  2003, The Messenger, 111, 36

\bibitem[{{Wilson} {et~al.}(2000){Wilson}, {Young}, \& {Shopbell}}]{WYS00}
{Wilson}, A.~S., {Young}, A.~J., \& {Shopbell}, P.~L. 2000, \apjl, 544, L27

\bibitem[{{Zanni} {et~al.}(2003){Zanni}, {Bodo}, {Rossi}, {Massaglia},
  {Durbala}, \& {Ferrari}}]{Zanea03}
{Zanni}, C., {Bodo}, G., {Rossi}, S., {Massaglia}, S., {Durbala}, A., \&
  {Ferrari}, A. 2003, \aap, in press

\bibitem[{{Ziegler} \& {Yorke}(1997)}]{ZY97}
{Ziegler}, U. \& {Yorke}, H.~W. 1997, Computer Physics Communications, 101, 54

\end{thebibliography}
%


\printindex
\end{document}